\documentclass[12pt]{article}
\usepackage{epsfig}

\newcommand{\be}{\begin {equation}}
\newcommand{\ee}{\end {equation}}
\newcommand{\betaph}{\beta_{\rm ph}}

\title{The Stern-Gerlach interaction between a traveling particle and a time varying magnetic field}
\author{\small\bf M. Conte$^1$, M. Ferro$^1$, G. Gemme$^1$, W.W. MacKay$^2$, R. Parodi$^1$ and 
   M. Pusterla$^3$\\
{\small\it ${}^{1)}$ Dipartimento di Fisica dell'Universit\`{a} di Genova and} \\ {\small\it INFN Sezione di 
Genova, Via Dodecaneso 33, 16146 Genova, Italy} \\
{\small\it ${}^{2)}$ C-A Dept., Brookhaven National Laboratory, Upton,
 NY 11973, USA}\\
{\small\it ${}^{3)}$ Dipartimento di Fisica dell'Universit\`{a} di Padova and} \\ {\small\it INFN Sezione di 
Padova, Via Marzolo 8, 35131 Padova, Italy}
} 
\date{}

\begin{document}

\maketitle

\begin{abstract}
The general expression of the Stern-Gerlach force is deduced for a charged
particle, endowed with a magnetic moment, which travels inside a time varying
magnetic field. Then, the energy integral of the Stern-Gerlach force is
evaluated in the case of a particle crossing a TE rf cavity with its magnetic 
moment oriented in different ways with respect as the cavity axis. We shall 
demonstrate that appropriate choices of the cavity characteristics and of the 
spin orientation confirm the possibility of separating in energy the opposite 
spin states of a fermion beam circulating in a storage ring and, in addition, 
make feasible an absolute polarimeter provide that a parametric converter 
acting between two coupled cavities is implemented.
\end{abstract}

  \vskip 1cm
  \begin{flushleft}
      {\small Report no.: INFN/TC-00/03, March, 22, 2000}  
  \end{flushleft}

\section{Introduction}
The Stern-Gerlach force acts on particles, carrying a magnetic
moment, which cross inhomogeneous magnetic fields. In a reference frame 
where particles are at rest, the expression of this force is
\be \vec f_{SG} = - {\nabla} U \label {fSG} \ee
where 
\be U = -\vec \mu \cdot \vec B \label {Energy} \ee
is the magnetic potential energy, and
\be \vec \mu = g{e\over 2m} \vec S \label {mu} \ee
is the magnetic moment. Here $e=\pm1.602\times 10^{-19}~\rm C$ is the 
elementary charge with $+$ for $p,e^+$ and $-$ for $\bar{p},e^-$, making 
$\vec \mu$ and $\vec S$ either parallel or antiparallel, respectively.
The rest mass, $m$, is $1.67\times 10^{-27}~{\rm kg}$ for 
$p,\bar p$ and $9.11\times 10^{-31}~{\rm kg}$
for $e^\pm$, and the relation between the 
gyromagnetic ratio $g$ and the anomaly $a$ is 

\be a = {{g-2}\over 2} = \cases{1.793 ~~~~~~~~~~~ (g=5.586) ~~~ 
{\mathrm {for}}~p,\bar p \cr 
1.160\times 10^{-3} ~ (g=2.002) {\mathrm ~~~ {for}}~e^\pm \cr} \label {ag} \ee 

In the rest system, the quantum vector $\vec S$, named spin, has modulus 
$|\vec S|=\sqrt{s(s+1)}\,{\hbar}$, and its component parallel to the magnetic 
field lines can take only the following values:

\be S_m=(-s,~-s+1,....,s-1,~s)\hbar, \label {S_main} \ee
where $\hbar=1.05\times 10^{-34} ~ \rm Js$ the reduced Planck's constant. 
Combining Eqs. (\ref{mu}) and (\ref{S_main}) we obtain for a generic 
spin-{$1\over 2$} fermion

\be \mu = |\vec \mu|=g{|e|\hbar\over 4m} \label {mumod} \ee 
or
\be \mu = \cases{1.41\times 10^{-26} ~\rm JT^{-1} \cr
9.28\times 10^{-24} ~\rm JT^{-1} \cr} \label {spinpe} \ee
Take note that the Bohr magneton is
\be \mu_B = 2\,[\mu/g]_{\mathrm {electron}} = 
9.27\times 10^{-24} ~\rm JT^{-1} \label {Bohrm} \ee

  Aiming to have the expression of the Stern-Gerlach force in the laboratory
frame, we have first to carry out the Lorentz transformation of the electric
and magnetic field from the laboratory frame, where we are at rest, to the
center-of-mass frame, where particles are at rest and we can correctly evaluate
such a force.
Then this force must be boosted back to the laboratory frame. 
All of these rather cumbersome operations will be discussed in the
next Section.
\section{Lorentz Boost of a Force}
In order to accomplish the sequence of Lorentz boosts
more easily, we choose a Cartesian 4-dimensional Minkowski metric 
\cite{Synge} $(x_1,x_2,x_3,x_4) = (x,y,z,ict)$, where $i=\sqrt{-1}$.
Therefore, the back-and-forth Lorentz transformations between
laboratory frame and particle's rest frame (usually labeled with a prime)
are the following:

\be \left(\matrix{x'\cr y'\cr z'\cr ict'\cr}\right) = 
M\,\left(\matrix{x\cr y\cr z\cr ict\cr}\right) = 
\left(\matrix{1 & 0 & 0 & 0\cr 0 & 1 & 0 & 0 \cr
0 & 0 & \gamma & i\beta\gamma \cr
0 & 0 & -i\beta\gamma & \gamma \cr}\right) 
\left(\matrix{x\cr y\cr z\cr ict\cr}\right) \Rightarrow 
\cases{x'=x\cr y'=y\cr z'=\gamma(z-\beta ct)\cr 
t' = \gamma\left(t-{\beta\over c}z\right) \cr} \label {LLR} \ee
$$ \left\{\beta = |\vec \beta|={|\vec v|\over c}, 
\gamma = {1\over \sqrt{1-\beta^2}}\right\} $$
and
\be \left(\matrix{x\cr y\cr z\cr ict\cr}\right) = 
M^{-1}\left(\matrix{x'\cr y'\cr z'\cr ict'\cr}\right) = 
\left(\matrix{1 & 0 & 0 & 0 \cr 0 & 1 & 0 & 0 \cr 
0 & 0 & \gamma & -i\beta\gamma \cr
0 & 0 & i\beta\gamma & \gamma \cr}\right) 
\left(\matrix{x'\cr y'\cr z'\cr ict'\cr}\right) \Rightarrow 
\cases{x=x'\cr y=y'\cr z=\gamma(z'+\beta ct')\cr 
t = \gamma\left(t'+{\beta\over c}z'\right) \cr} \label {LRL} \ee
Moreover, combining both eqs. (\ref{LLR}) and (\ref{LRL}), we obtain the 
following expressions for the partial derivatives:

\be {\partial\phantom{x'} \over \partial x'}
 = {\partial\phantom{x} \over \partial x},\quad
{\partial\phantom{y'} \over \partial y'}
 = {\partial\phantom{y} \over \partial y} \label {dxdy} \ee
\be {\partial\phantom{z'} \over \partial z'}
  = \gamma\,\left(
       {\partial\phantom{z} \over \partial z} + {\beta\over c}
       {\partial\phantom{t} \over \partial t}\right) \label {dz} \ee

  The 4-vector formalism is still applied for undergoing the Lorentz 
transformation of a force. First of all, let us define as 4-velocity the
quantity

\be u_\mu = {dx_\mu\over d\tau} \label {4v} \ee
where
\be d\tau = {ds\over c} = {dt\over \gamma} \label {propt} \ee
is the differential of the proper time.
We define the 4-momentum as the product of the rest mass $m$ times
the 4-velocity, i.e.

\be P_\mu = m\,u_\mu = (\vec p,i\gamma mc) \label {4mom} \ee
The 4-force is the derivative of the 4-momentum (\ref{4mom})
with respect to the proper time, that is

\be F_\mu = {dP_\mu\over d\tau} = \left(\gamma{d\vec p\over dt},
i{\gamma\over c}{d(\gamma mc^2)\over dt}\right) = \left(\gamma\vec f,
i{\gamma\over c}{dE_{\mathrm {tot}}\over dt}\right) \label {4f} \ee
where $\vec f$ is the ordinary force. In the c.m. system eq. (\ref{4f})
reduces to

\be F'_\mu = (\vec f',0) \label {4fR} \ee
since $\gamma'=1$ and $E'_{\mathrm {tot}}=mc^2$ is a constant. Bearing in
mind the last step of the whole procedure, i.e. the boost of any force from 
rest to laboratory frame, we have to use the relation

\be F_\mu = M^{-1} F'_\mu =
\left(\matrix{\gamma f_x \cr \gamma f_y \cr \gamma f_z \cr F_4 \cr}
\right) = 
\left(\matrix{1 & 0 & 0 & 0 \cr 0 & 1 & 0 & 0 \cr 
0 & 0 & \gamma & -i\beta\gamma \cr
0 & 0 & i\beta\gamma & \gamma\cr}\right) 
\left(\matrix{f'_x\cr f'_y\cr f'_z\cr 0\cr}\right) =
\left(\matrix{f'_x\cr f'_y\cr \gamma f'_z\cr 
i\beta\gamma f'_z \cr}\right) \label {4fL} \ee
or

\be \vec f_\perp = {1\over \gamma}\,\vec f'_\perp \label {fperp} \ee
\be \vec f_\parallel = \vec f'_\parallel ~~~ (f_z = f'_z) \label {fpar} \ee
\section{Stern-Gerlach Force}
The Stern-Gerlach force, as described by eq. (\ref{fSG}), must be
evaluated in the particle rest frame where it takes the form

\be \vec f'_{SG} = \nabla'(\vec \mu^* \cdot \vec B') =
{\partial \over \partial x'} (\vec \mu^* \cdot \vec B') {\hat {x}} +
{\partial \over \partial y'} (\vec \mu^* \cdot \vec B') {\hat {y}}  +
{\partial \over \partial z'} (\vec \mu^* \cdot \vec B') {\hat {z}} 
\label {f'SG} \ee
having defined the magnetic moment as $\mu^*$, rather than $\mu'$, for
opportune reasons. By applying the transformations (\ref{dxdy}), 
(\ref{fperp}) and (\ref{fpar}), the force ({\ref{f'SG}}) is boosted 
to the laboratory system becoming
\be \vec f_{SG} = 
{1\over \gamma} {\partial \over \partial x} (\vec \mu^* \cdot \vec B') 
{\hat {x}} +
{1\over \gamma} {\partial \over \partial y} (\vec \mu^* \cdot \vec B')
{\hat {y}} +
{\partial \over \partial z'} (\vec \mu^* \cdot \vec B') {\hat {z}} 
\label {f-SG} \ee
Bearing in mind the Lorentz transformation \cite{Jack} 
of the fields $\vec E,\vec B$ and $\vec E',\vec B'$ 

\be \vec E' = \gamma(\vec E + c\vec \beta\times\vec B) -
{\gamma^2\over \gamma+1}\vec \beta(\vec \beta\cdot\vec E) \label {ER} \ee
\be \vec B' = \gamma\left(\vec B - {\vec \beta\over c}\times\vec E\right) -
{\gamma^2\over \gamma+1}\vec \beta(\vec \beta\cdot\vec B) \label {BR} \ee
the energy $(\vec \mu^* \cdot \vec B')=\mu_xB'_x + \mu_yB'_y + \mu_zB'_z$ 
becomes

\be (\vec \mu^* \cdot \vec B') = \gamma\mu^*_x
\left(B_x + {\beta\over c}E_y\right) + \gamma\mu^*_y
\left(B_y - {\beta\over c}E_x\right) + \mu^*_zB_z \label {magen} \ee

  If we introduce eq. (\ref{magen}) into eq. (\ref{f-SG}) and take 
into account eq. (\ref{dz}), we can finally obtain the Stern-Gerlach 
force components in the laboratory frame:

\be f_x = \mu^*_x\left({\partial B_x\over \partial x} + {\beta\over c}
                       {\partial E_y\over \partial x}\right) +
          \mu^*_y\left({\partial B_y\over \partial x} - {\beta\over c}
                       {\partial E_x\over \partial x}\right) +
 {1\over \gamma}\mu^*_z{\partial B_z\over \partial x}     \label {fSGx} \ee
\be f_y = \mu^*_x\left({\partial B_x\over \partial y} + {\beta\over c}
                       {\partial E_y\over \partial y}\right) +
          \mu^*_y\left({\partial B_y\over \partial y} - {\beta\over c}
                       {\partial E_x\over \partial y}\right) +
 {1\over \gamma}\mu^*_z{\partial B_z\over \partial y}     \label {fSGy} \ee
\be f_z = \mu^*_xC_{zx} + \mu^*_yC_{zy} + \mu^*_zC_{zz}    \label {fSGz} \ee
with
\be C_{zx} = \gamma^2\left[\left({\partial B_x\over \partial z} + 
                  {\beta\over c} {\partial B_x\over \partial t}\right) +
{\beta\over c}\left({\partial E_y\over \partial z} + {\beta\over c}
                           {\partial E_y\over \partial t}\right)\right] 
                                                \label {Czx} \ee
\be C_{zy} = \gamma^2\left[\left({\partial B_y\over \partial z} + 
                  {\beta\over c} {\partial B_y\over \partial t}\right) -
{\beta\over c}\left({\partial E_x\over \partial z} + {\beta\over c}
                           {\partial E_x\over \partial t}\right)\right] 
                                                \label {Czy} \ee
\be C_{zz} = \gamma\left({\partial B_z\over \partial z} + 
{\beta\over c} {\partial B_z\over \partial t}\right) \label {Czz} \ee

\section{The Rectangular Cavity}
\begin{figure}[t]
\begin{center} \mbox{\epsfig{file=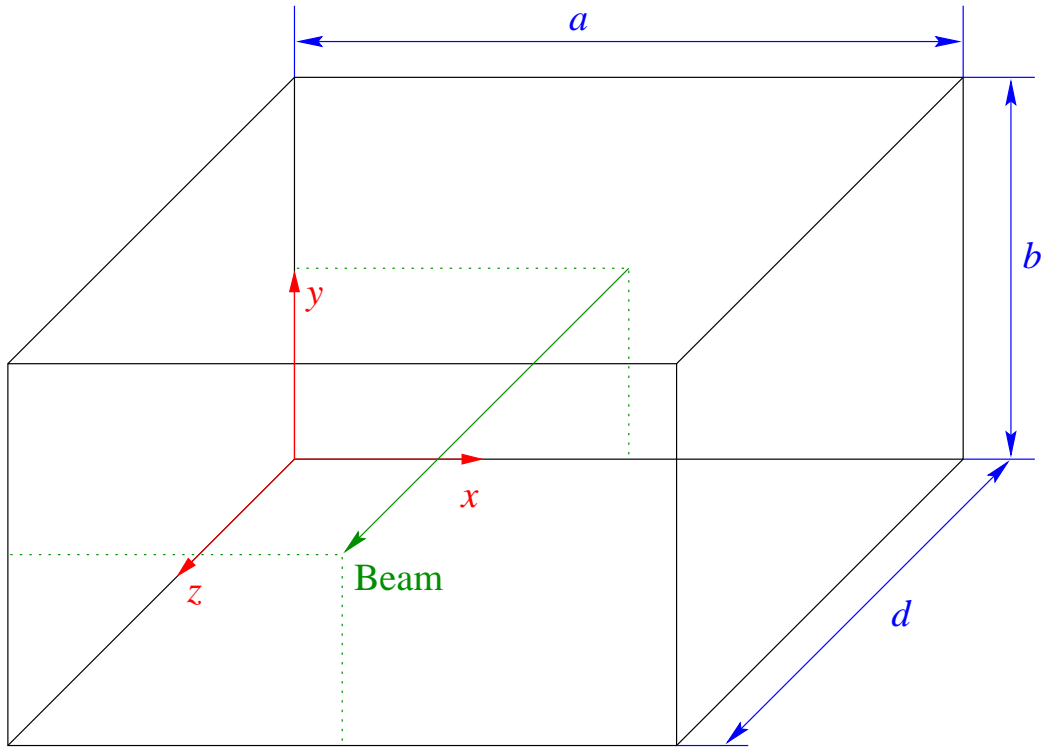}} \end{center}
\caption{Sketch of  the rectangular cavity; take note that coordinates of the beam axis are x=a/2 and y=b/2.}
\label{cavity}
\end{figure}

In order to simplify our calculations without loosing the general 
physical meaning, we shall consider a rectangular resonator, as the one 
shown in Fig.\ref{cavity}, which is characterized \cite{Ramo} by the following field 
components:

\be B_x = 
- {B_0\over K_c^2} \left({m\pi\over a}\right) \left({p\pi\over d}\right)
 \sin\left({m\pi x\over a}\right) \cos\left({n\pi y\over b}\right) 
 \cos\left({p\pi z\over d}\right) \cos\,\omega t       \label {Bx} \ee
\be B_y = 
- {B_0\over K_c^2} \left({n\pi\over b}\right) \left({p\pi\over d}\right)
 \cos\left({m\pi x\over a}\right) \sin\left({n\pi y\over b}\right) 
 \cos\left({p\pi z\over d}\right) \cos\,\omega t       \label {By} \ee
\be B_z = B_0 \cos\left({m\pi x\over a}\right) \cos\left({n\pi y\over b}\right) 
 \sin\left({p\pi z\over d}\right) \cos\,\omega t       \label {Bz} \ee
\be E_x =
- B_0 \left({n\pi\over b}\right) {\omega\over K_c^2} 
 \cos\left({m\pi x\over a}\right) \sin\left({n\pi y\over b}\right) 
 \sin\left({p\pi z\over d}\right) \sin\,\omega t       \label {Ex} \ee 
\be E_y =
  B_0 \left({n\pi\over b}\right) {\omega\over K_c^2} 
 \sin\left({m\pi x\over a}\right) \cos\left({n\pi y\over b}\right) 
 \sin\left({p\pi z\over d}\right) \sin\,\omega t       \label {Ey} \ee 
\be E_z = 0 ~~~ ({\mathrm {as~typical~for~a~TE~mode}})  \label {Ez} \ee
where $B_0$ is the amplitude of the $B_z$-component and

\be K_c = \sqrt{\left({m\pi\over a}\right)^2 + \left({n\pi\over b}\right)^2}
                                                        \label {kc} \ee
\be {\omega\over c} = K = {2\pi\over \lambda} = 
\sqrt{\left({m\pi\over a}\right)^2 + \left({n\pi\over b}\right)^2 +
      \left({p\pi\over d}\right)^2}                  \label {omega} \ee
The wave's phase velocity is $v_{\rm ph}=\betaph c$
where
\be \betaph = {K\over \sqrt{K^2-K_c^2}} =
\sqrt{1+\left({md\over pa}\right)^2+\left({nd\over pb}\right)^2} \label {wpv}
\ee
                 
   We have to recall that the polarization of a beam, revolving in a ring
whose guide field is $\vec B_{\mathrm {ring}}$, can be defined as

\be P = {{N_\uparrow - N_\downarrow}\over {N_\uparrow + N_\downarrow}}
        \label {Pol} \ee
where
\break\noindent
$N_\uparrow$ = No. Particles Spin Up (e.g. parallel to 
$\vec B_{\mathrm {ring}}$)\\
\noindent
$N_\downarrow$ = No. Particles Spin Down (antiparallel to 
$\vec B_{\mathrm {ring}}$)
\par\vspace{0.2 true cm}\noindent
and $P$ indicates the macroscopic average over the particle distribution in 
the beam, which is equivalent to the quantum mechanical expectation value 
found by means of the quantum statistical matrix. Obviously, an unpolarized 
beam has $P=0$ or $N_\uparrow$ = $N_\downarrow$. 

  A quick comparison among the SG-force components, given by the set of 
equations (\ref{fSGx})-(\ref{Czz}), suggests that $f_z$ will dominate
at high energy, since it contains terms proportional to $\gamma^2$,
whereas the transverse components have terms independent of $\gamma$, not
to mention the $\gamma^{-1}$ terms.

  The most appropriate choice of the spin orientation seems to be the one 
parallel to ${\hat {y}}$ i.e. to $\vec B_{\mathrm {ring}}$, i.e. the force
component is the one given by eq. (\ref{fSGz}) with the insertion of 
eq. (\ref{Czy}). This means that particles undergoing energy gain (or loss) 
don't need any spin rotation while entering and leaving the rf cavity, beyond 
the advantage of having to deal with a force component proportional to 
$\gamma^2$. Choosing the simplest ${\mathrm {TE}}_{011}$ mode, the quantities
(\ref{kc}), (\ref{omega}) and (\ref{wpv}) reduce to
\be k_c = {\pi\over b} \label {kc011} \ee
\be \omega = c \sqrt{\left({\pi\over b}\right)^2 +
                     \left({\pi\over d}\right)^2}      \label {om011} \ee
\be \betaph = 
\sqrt{1+\left({d\over b}\right)^2}
      \label {wpv011} \ee
Setting $x={a\over 2}$ and $y={b\over 2}$ the field components
along the beam axis become

\be B_x = B_z = 0 \label {Bxz0} \ee
\be B_y = - B_0 {b\over d}\, \cos\left({\pi z\over d}\right) \cos\,\omega t  
                                                        \label {By011} \ee
\be E_x = - \omega\,B_0 {b\over \pi}\, \sin\left({\pi z\over d}\right) 
                                       \sin\,\omega t   \label {Ex011} \ee
\be E_y = E_z = 0 \label {Eyz0} \ee
therefore the force component $f_z$ can be written as

\be f_z = \mu^*\gamma^2 B_0 b\,\left\{ {1\over \pi}
          \left[\left({\pi\over d}\right)^2 + 
          \left({\beta\omega\over c}\right)^2\right]
          \sin\left({\pi z\over d}\right) \cos\,\omega t +
          {2\over d} \left({\beta\omega\over c}\right)
          \cos\left({\pi z\over d}\right) \sin\,\omega t\right\} 
                                                         \label {fz011} \ee

  For completeness, we shall also analyze the possibility of using 
a spin orientation parallel to ${\hat {z}}$, i.e. to the motion direction,
even though this option requires a system of spin rotators and looses
a factor of $\gamma$ in the force component.
\section{Involved Energy}
The energy gained, or lost, by a particle with a magnetic moment after
having crossed a rf cavity can be evaluated by integrating the Stern-Gerlach 
force (\ref{f-SG}) over the cavity length, namely:
\be \Delta U = \int_{0}^{d} dU = \int_{0}^{d} \vec f \cdot d\vec r = 
          \int_{0}^{d} f_z dz = 
          \int_{0}^{d} \mu^*C_{zy}\,dz \label {DUByz} \ee
Bearing in mind eq. (\ref{fz011}) and carrying out the trivial substitution 
$\omega t={\omega z\over \beta c}$, the integral (\ref{DUByz}) becomes

$$ \Delta U = \mu^*\gamma^2 B_0 b\,\left\{ {1\over \pi}
          \left[\left({\pi\over d}\right)^2 + 
          \left({\beta\omega\over c}\right)^2\right] I_1 +
          {2\over d} \left({\beta\omega\over c}\right) I_2\right\} $$
with

$$ I_1 = \int_{0}^{d} \sin\left({\pi z\over d}\right) 
        \cos\left({\omega z\over \beta c}\right) dz = {{\pi\over d}\over 
     \left({\pi\over d}\right)^2 - \left({\omega \over \beta c}\right)^2}
     \left[1+\cos\left({\omega d\over \beta c}\right)\right] $$    
$$ I_2 = \int_{0}^{d} \cos\left({\pi z\over d}\right) 
        \sin\left({\omega z\over \beta c}\right) dz = 
       - {{\omega\over \beta c}\over 
     \left({\pi\over d}\right)^2 - \left({\omega \over \beta c}\right)^2}
     \left[1+\cos\left({\omega d\over \beta c}\right)\right] $$    
or

\be \Delta U = \mu^*\gamma^2 B_0\, {b\over d} \,
{\left({\pi\over d}\right)^2
    + \left({\beta\omega \over c}\right)^2
    - 2\left({\omega \over c}\right)^2
\over
\left({\pi\over d}\right)^2 - \left({\omega \over \beta c}\right)^2
}
\left[1+\cos\left({\omega d\over \beta c}\right)\right]
\label {dU-apr} \ee

Taking into account the stationary wave conditions
(eqs. \ref{om011} and \ref{wpv011})
pertaining to the $\rm TE_{011}$ mode,
the length of the cavity can be expressed as
\be d = {1\over 2}\betaph \lambda \label {cavl} \ee
which allows us
to write eq. (\ref{dU-apr}) as 
\be
\Delta U = \gamma^2\beta^2\mu^*B_0\, {b\over d}\,
{1+\betaph^2(\beta^2-2)\over \beta^2-\betaph^2}
              \left(1+\cos{\betaph\over \beta}\pi\right)
\ee
In the ultrarelativistic limit ($\gamma \gg 1$ and $\beta \simeq 1$),
\be \Delta U \simeq \mu^*B_0\, {b\over d}\, \gamma^2 (1+\cos\betaph\pi) =
                  2\,\mu^*B_0\, {b\over d}\, \gamma^2 ~~~ 
                   (\betaph = {\hbox{even integer}})  \label {dUf} \ee

   As hinted before, let us evaluate the work-energy integral when
the particle enters into the cavity with its spin parallel to $\hat {z}$.
In this example we must choose the mode ${\mathrm {TE}}_{021}$ as the 
lowest one; then we have from eqs. (\ref{Bz}) and (\ref{Czz}) respectively

\be B_z = - B_0\,\sin\left({\pi z\over d}\right)\,\cos\,\omega t 
                                                       \label {Bzz} \ee
\be f_z = \mu^*C_{zz} = - \mu^*\gamma B_0 
  \left[{\pi\over d}\cos\left({\pi z\over d}\right) \cos\,\omega t -
  \left({\beta\omega\over c}\right)
        \sin\left({\pi z\over d}\right) \cos\,\omega t\right]
                                                         \label {fzz} \ee
and proceeding as above we obtain

\be \Delta U = \mu^*B_0 \gamma\, {\pi\over d} \,
  {{\omega\over \beta c} - {\beta c\over \omega}\over
  \left({\pi\over d}\right)^2 - \left({\omega\over \beta c}\right)^2}
  \sin\left({\omega d\over \beta c}\right)  \label {dUz} \ee 
and
\be \Delta U = {\mu^*B_0\over \gamma}{\betaph\beta\over \betaph^2-\beta^2}
               \sin\left({\betaph\over \beta}\pi\right)  \label {dUzz} \ee    
or ultrarelativistically
\be \Delta U \simeq {\mu^*B_0\over \gamma}{\betaph\over \betaph^2-1}
                    \sin\betaph\pi,
\qquad
    \Delta U_{\rm max} \sim - 1.62 {\mu^*B_0\over \gamma} \quad
                                          (\hbox{when $\betaph \sim 1.13$})
  \label {dUzzz} \ee    
confirming a result \cite{Waldo} already achieved.

   Before making up our mind, we need to compare the energy gain/loss due
to the Stern-Gerlach interaction with the same quantity caused by the electric
field. To this aim, we emphasize that

\be dU_E = \vec f_E \cdot d\vec r = eE_x dx           \label {dUE*1} \ee
as can be easily understood looking at eqs. (\ref{Ex011}) and (\ref{Eyz0}). 
Since the carrier particle travels from 0 to $d$ along the $z$-axis, the only
integral which makes sense is the following:

\be \Delta U_E = \int_{0}^{d} eE_x\,dx = \int_{0}^{d} eE_x\,{dx\over dz}dz =
              \int_{0}^{d} eE_x\,x'dz \label {dUE*2} \ee
or
$$ \Delta U_E = - x'e\omega B_0 {b\over \pi}\int_{0}^{d}
               \sin\left({\pi z\over d}\right)
               \sin\left({\omega z\over \beta c}\right) dz =
               - x'e\omega B_0 {b\over d}
     {\sin\left({\omega d\over \beta c}\right)\over
     \left({\pi\over d}\right)^2 - \left({\omega\over \beta c}\right)^2} $$
or
\begin{equation}
\Delta U_E = \left [ e \omega B_0 {bd\over \pi^2}
  {\beta^2\over \betaph^2 - \beta^2}\sin{\betaph\over \beta}\pi \right ] x' = \kappa x' 
\label {dUE*3}
\end{equation}
having proceeded as before.

  We recall that the Stern-Gerlach interaction in the realm of particle
accelerators has been proposed either for separating in energy particles 
with opposite spin states, the well known \cite{CoPePu} spin-splitter 
concept, or for settling an absolute polarimeter \cite{Pete}.

  As far as the spin-splitter is concerned, we quickly recall that spin up 
particles receive (or loose) that amount of energy given by eq. (\ref{dUf}) 
at each rf cavity crossing, and this will take place all over the time 
required. Simultaneously, spin down particles behave exactly in the opposite 
way, i.e. they loose (or gain) the same amount of energy turn after turn.
The actual most important issue is that the energy exchanges sum up coherently.
More quantitatively, we may indicate as the final energy separation after
$N$ revolutions:

\be \Delta_{\uparrow\downarrow} = \sum\,\{\Delta_\uparrow -
         (-\Delta_\downarrow)\} = 4{b\over d} N\,\mu^*B_0\,\gamma^2 \simeq
         4\,N\,\mu^*B_0\,\gamma^2  \label {En-Sep} \ee
Instead, the adding up of the energy contribution (\ref{dUE*3}) due to
the electric field is

\be (\Delta U_E)_{tot} = \sum\,\Delta U_E = \kappa\,\sum\,x' = 0 
                                                   \label {betav} \ee
since $x'$ changes continuously its sign with a periodicity related to
the period of the betatron oscillations.

The result (\ref{En-Sep}), together with the demonstration
(\ref{betav}), would seem to provide very good news for the spin-splitter
method!

    As far as the polarimeter is concerned, we have to bear in mind that
we are interested in the instantaneous interaction between magnetic moment
and the rf fields: therefore the zero-averaging due to the incoherence of
the betatron oscillations would not help us. Notwithstanding, if we set
$\betaph$ equal to an {\it integer} in eq. (\ref{dUE*3}), we have for 
U.R. particles:
\be
\Delta U_E = {x'e \omega B_0\,bd\over \pi^2(\betaph^2-1)}
  \sin\left(\betaph\pi + {\betaph\pi\over 2\gamma^2}\right) \simeq 
  \pm {x'bd\over 2\pi} {\betaph\over \betaph^2-1} {e\omega B_0\over \gamma^2}
\label {dUE*4}
\ee
Then this $1/\gamma^2$ dependence of the spurious signal, compared
to the $\gamma^2$ dependence of the signal (\ref{dUf}) to be measured, 
sounds interesting for the feasibility of this kind of polarimeter;
however, one must realize that if $\betaph$ is not exactly an integer, then
eq.~(\ref{dUE*4}) would become
\be
\Delta U_E \sim
  \pm {x'bd\over 2\pi}
 {e\omega B_0\over \betaph^2-1}
\left(\epsilon + {\betaph\over \gamma^2}\right)
\label {dUE*5}
\ee
where $\epsilon$ is the error in $\betaph$.
\section{A Few Numerical Examples}
The spin-splitter principle requires a repetitive crossing of
$N_{\rm cav}$
cavities distributed along the ring, each of them resonating in the
TE mode. After each revolution, the particle experiences a variation, or 
{\it kick}, of its energy or of its momentum spread 

\be \zeta = {\delta p\over p} = {1\over \beta^2} {\delta E\over E} \simeq
            {N_{\rm cav}\Delta U\over E} \simeq 
            {2\sqrt{3}\over 3}\,N_{\rm cav}{B_0\over B_\infty}\gamma
\label {k1}
\ee
having made use of eq. (\ref{dUf}), further simplified by reasonably setting 
$\betaph=2$, and with 

\be B_\infty = {mc^2\over \mu^*} = {1.503 \times 10^{-10} ~{\rm J}\over
     1.41 \times 10^{-26} ~{\rm JT}^{-1}} \simeq 10^{16}~{\rm T}  \label{Binf} \ee
for (anti)protons. From eq. (\ref{k1}) we may find as the number of turns 
needed for attaining a momentum separation equal to 
2$\left({\Delta p\over p}\right)$ 

\be N_{\rm SS} = {\left({\Delta p\over p}\right)\over \zeta} = 
                  {\sqrt{3}\over 2\,N_{\rm cav}\gamma} {B_\infty \over B_0}
                      \left({\Delta p\over p}\right)  \label {LSS} \ee
Multiplying $N_{\rm SS}$ by the revolution period $\tau_{\rm rev}$ we obtain

\be \Delta t = N_{\rm SS}\tau_{\rm rev} \label {D-time} \ee
as the actual time spent in this operation. For the sake of having 
some data, we consider RHIC \cite{RHIC} and HERA \cite{HERA} whose essential
parameters are shown in Table I together with what can be found by 
making use of eqs. (\ref{LSS}) and (\ref{D-time}) where $B_0\simeq 0.1~T$ and 
$N_{\rm cav}=200$ are chosen as realistic values.
\pagebreak
\begin {center}
{\bf Table I: RHIC and HERA parameters}
\par\vspace {0.5 cm}
\begin{tabular}{|c|c|c|}
\hline
                    &  RHIC               &  HERA                  \\
\hline
 E(GeV)             &   250               &  820                   \\
\hline
 $\gamma$           & 266.5               & 874.2                  \\
\hline
$\tau_{\rm rev}(\mu\rm s)$ &  12.8               &  21.1                  \\
\hline
${\Delta p\over p}$ & $4.1\times 10^{-3}$ & $5\times 10^{-5}$      \\
\hline
$N_{\rm SS}$            & $6.67 \times 10^9$    & $2.48 \times 10^7$   \\
\hline
$\Delta t$          & $8.52 \times 10^4 ~ s\,\simeq\,23.7 ~ h$ & 523 ~ s \\
\hline
\end{tabular}
\end {center}
\par\vspace {0.5 cm}

   In the example of the polarimeter we have to pick up a signal generated
at each cavity crossing. Therefore, making use of eq. (\ref{dUf}) we have for
a bunch train made up of $N$ particles the total energy transfer
\be
\Delta U \approx 2 N P \mu^* B_0 \, \frac{b}{d}\, \gamma^2   \label {gainpol}
\ee
where $P$ is the beam polarization slightly modified with respect the
definition (\ref{Pol})

\be P = {{N_\rightarrow - N_\leftarrow}\over {N_\rightarrow + N_\leftarrow}}
        \label {Polmod} \ee
The average power transferred will be
\be
W = \frac{\Delta U}{\tau_{\rm rev}} \label {powerpol} \ee

   If we operate our cavity as a parametric converter 
\cite{ManRow}\cite{Louis}, with an initially empty level, we have for the power 
transferred to this empty level
\be
W_2 = \frac{\omega_{\rm rf}}{\omega_{\rm rev}} \, W =
      \frac{\nu_{\rm rf}}{\nu_{\rm rev}} \, W \label {fracpow1}
\ee 
where $\nu_{\rm rf}$ is the working frequency of the resonant cavity 
(typically in the GHz range), and $\nu_{\rm rev}$ is the revolution frequency.
Putting all together we have 
\be
W_2 \simeq 2 \, P \, \frac{\nu_{\rm rf}}{\nu_{\rm rev}} \, \mu^* B_0 \, 
\frac{b}{d} \, \gamma^2 \label {fracpow2}
\ee

   A feasibility test of the polarimeter principle has been proposed 
\cite{Pete} and studied \cite{Iron} to be carried out in the 500~MeV
electron ring \cite{Bates} of MIT- Bates, whose main characteristics are

\par\vspace {1 cm}
\begin {center}
{\bf Table II: MIT-Bates parameters}
\par\vspace {0.5 cm}
\begin{tabular}{|c|c|}
\hline
$\tau_{\rm rev}$	&	634 nsec	\\
\hline
$\nu_{\rm rev}$	&	1.576 MHz	\\
\hline
$N_{electrons}$	&	$3.6 \times 10^8 \cdot 225 = 8.1 \times 10^{10}$ \\
\hline
$\gamma$	& $\simeq 10^3$ \\
\hline
$b/d$	&	$\sqrt{3}/3$	\\
\hline
$B_0$	&	$\simeq 0.1$ T \\
\hline
$\nu_{\rm rf} / \nu_{\rm rev}$ & $\approx 10^3$ \\
\hline
$\mu^*$ & $9.27 \times 10^{-24} ~{\rm JT}^{-1}$ \\
\hline
\end{tabular}
\end {center}

\par\vspace {0.5 cm}
\noindent
and, since polarized electrons can be injected into this ring but 
precessing on a horizontal plane, the ${\mathrm {TE}}_{101}$ mode is
more appropriate than the ${\mathrm {TE}}_{011}$ as we shall have to use
$B_x$ rather than $B_y$: a choice that does not make any substantial
difference! From the above data we obtain

\be W_2 \simeq 137  P \,{\rm watts} \label {pow2} \ee

   Paradoxically, even for an almost unpolarized beam with $N_\rightarrow - 
N_\leftarrow = 1$ and, as a consequence of eq. (\ref{Polmod}), with
$P \simeq 1.23 \times 10^{-11}$, we should obtain 
$W_2 \approx 1.7 ~{\rm nW}$, which can be easily measured.

   As a last check, let us compare the energy exchanges 
($\vec \mu \Leftrightarrow \vec B$) and ($e \Leftrightarrow \vec E$). Taking 
into account eqs. (\ref{cavl}), (\ref{dUf}) and (\ref{dUE*4}), and setting 
$x' \simeq 1~\rm mrad$, $\betaph = 2$ and $\lambda = 10 ~\rm cm$,
we have for the Bates-MIT ring:

\be r = \frac {\Delta U_E} {\Delta U} = \frac {x'} {8} 
        \, \frac {\betaph^3} {\betaph^2-1} \, \frac {\lambda e c} {\mu^*}
        \, \frac {1} {\gamma^4} = 1.72 \times 10^{-4}     \label {check} \ee
i.e. the spurious signal, depending upon the electric interaction between
$e$ and $\vec E$, is absolutely negligible with respect the measurable 
signal generated by the magnetic interaction.
\section {Conclusions}
There is not too much to add to what has been found in the previous
Sections, aside from performing more accurate calculations and numerical
simulations. The Stern-Gerlach interaction seems very promising either for
attaining the self polarization of a $p(\bar{p})$ beam or for realizing an
absolute polarimeter.

   In the first example the problem raised \cite{CoMacPa} by the rf 
filamentation still holds on, although some tricks can be conceived: the
extreme one could be the implementation of a triangular waveform in the
TM cavity which bunches the beam.

   The second example requires nothing but to implement that experimental
test at the Bates-MIT electron ring.
\end{document}